# Quantum Technologies in the Telecommunications Industry.

Vicente Martin[*,1], Juan Pedro Brito[1], Carmen Escribano[1], Marco Menchetti[2], Catherine White[2], Andrew Lord[2], Felix Wissel[3], Matthias Gunkel[3], Paulette Gavignet[4], Naveena Genay[4], Olivier Le Moult[4], Carlos Abellán[5], Antonio Manzalini[6], Antonio Pastor-Perales[7], Victor López[7], Diego López[7].

[1]Center for Computational Simulation, DLSIIS/DMATIC, ETSI Informáticos, Universidad Politécnica de Madrid, Boadilla del Monte, 28660 Madrid. Spain

[2] British Telecom Research Laboratory, Adastral Park, Ipswich, IP5 3RE. UK

[3] Deutsche Telekom Technik GmbH, Heinrich-Hertz-Straße 3-7, 64295 Darmstadt. Germany

[4] Orange Labs, 2 Avenue Pierre Marzin, 22300 Lannion. France

[5] Quside Technologies, Esteve Terradas 1, Castelldefels, 08860 Barcelona. Spain

[6] Telecom Italia, Via Reiss Romoli 274, 10148 Turin. Italy.

[7]Telefónica gCTIO/I+D, Ronda de la Comunicación s/n, 28050 Madrid. Spain

**Abstract:**

*Quantum based technologies have been fundamental in our world. After producing the laser and the transistor, the devices that have shaped our modern information society, the possibilities enabled by the ability to create and manipulate individual quantum states opens the door to a second quantum revolution. In this paper we explore the possibilities that these new technologies bring to the Telecommunications industry.*

# Introduction

Quantum mechanics was a revolution in our understanding of the world. Since its early days in the 1920s it has not only contributed to our knowledge, but produced technological advances that completely changed our world and are at the basis of our society. The theory enables technologies such as: the transistor, the key component of our electronics and information science; the laser, at the base of our communications; simulations that have made possible our materials science, quantum chemistry and modern pharmacology. It is safe to say that a large amount of our economy is based on quantum technologies. However, just as Maxwell's equations of 1863 took a century to master and start flourishing in a widespread technology, it is also arguable that these applications of quantum mechanics are leading to a second quantum revolution that could shape, again, our society and economy. It took over half a century for quantum mechanics to start evolving from a purely physical theory, with implications even at the philosophical level, to a science with a deep impact on how we understand and process information; the basis of our information technology. Landauer's view "Information is physical" [1] summarizes a new vision in which information cannot be detached from its physical embodiment, changing the way of thinking from the purely mathematical construct of Shannon's bit to what could be really implemented in the physical world.

[*] Corresponding author: vicente.martin@upm.es, juanpedro.brito@upm.es, cescribano@fi.upm.es, marco.menchetti@bt.com, catherine.white@bt.com, andrew.lord@bt.com, Felix.Wissel@telekom.de, Matthias.Gunkel@telekom.de, paulette.gavignet@orange.com, naveena.genay@orange.com, olivier.lemoult@orange.com, cabellan@quside.com, antonio.manzalini@telecomitalia.it, antonio.pastorperales@telefonica.com, victor.lopezalvarez@telefonica.com, diego.r.lopez@telefonica.com

It was not until 1982 when quantum states were clearly shown to have information processing properties that were not envisioned in the predominant computing paradigm. The so-called "no-cloning theorem" [2] [3] made clear that, when information is encoded at the lowest physical level possible, new properties had to be considered. The pioneering work of Benioff and Feynman [4] [5] in the same year made also clear that quantum computing might be actually more powerful than classical machines. Others, like Wiesner [6], had already envisioned how these properties could be used advantageously in certain cases. In 1984, the first quantum cryptography protocol was published by Bennett and Brassard [7], allowing the creation of symmetric keys with perfect -in its mathematical embodiment- secrecy between the end points of a channel able to transmit quantum correlations. Its first, albeit primitive, implementation in 1989 [8] added credibility to its claim of being a useful technology and started in practice the field of Quantum Key Distribution. The implementation over a meagre free-space channel of just 25cm was just the first of a long series that have demonstrated the possibility to do QKD over several hundreds of km in optical fiber and thousands in free-space using satellites.
However, due to the same no-cloning principle that affords its security, the faithful amplification of quantum signals is not possible, thus ultimately limiting the maximum distance achievable. It is important to note that no-cloning does not preclude the existence of quantum repeaters, a device that is able to establish quantum correlations over unlimited distances without actually copying states, hence also able to free QKD from any distance or losses constraint. Quantum repeaters were described in 1998 [9].

1994 saw the publication of Shor's algorithm [10], able to solve the discrete logarithm and integer factoring problems. This brought quantum computing to the attention of a much broader community because of its implications in cryptanalysis. By showing how quantum computers could solve the factoring problem in polynomial time, a problem that resisted the efforts of mathematicians during centuries, Shor broke the algorithms at the basis of modern public-key cryptography. The threshold theorem [11] showing that, in principle, an unlimited computation can be performed using a quantum computer was demonstrated in 1998. The interest spurred by these and other results caused that by the early 2000 the kernels of the main algorithms were already demonstrated in primitive quantum processors.

Fast forward to the current day, and it has been demonstrated that there are problems that can be efficiently solved on a quantum computer that cannot be solved on a classical computer [12], quantum key distribution systems are already in a commercial stage and a large number of companies and governments have invested heavily in researching quantum information technologies, including also important aspects like sensing and metrology. Although there are still important issues to be solved for its widespread adoption, it seems unquestionable that quantum technologies are called to play a relevant role in the technological panorama of the next decades.

In the present paper, we will describe the main quantum technologies which are expected to play a role in the telecommunications industry. Quantum communications are possibly the most advanced and closest to market adoption of all quantum technologies. However, they are not the only ones, and we will also discuss aspects of quantum computing and quantum metrology that are expected to be applicable in the field of telecommunications. Whenever possible, we illustrate their applications using real-world use-cases. The paper is thus divided in three sections; first and foremost, with short-term applicability, is quantum communications, especially QKD [13] [14, 15], and quantum random number generation, followed by computing and metrology (timing). Quantum computing, given the state of the art, is necessarily more speculative, but potentially very important and is advancing quickly. Albeit still a little bit controversial, a quantum computer with 53 qubits has performed in minutes calculations that would have taken thousands of years in a classical computer [16, 17, 18]. Each section has an introduction to the technology followed by its application and use-cases. The paper finishes with some concluding remarks.

## Quantum Communications

Of all quantum information technologies, quantum communication is possibly the most advanced one. Its basic purpose is to transmit quantum signals from an emitter to a receiver. This apparently simple task is in fact very complicated due to the same properties of quantum signals that make them so interesting and capable of performing tasks that are impossible when using classical signals alone. In this particular context, the most important one is the impossibility of faithfully copying unknown quantum signals. This makes impossible to amplify them, thus limiting the reach of quantum communications by directly sending quantum signals through an absorbing medium[†]. In practice, when using optical fiber in the most transparent window, where optical losses are around 0.2 dB/km in the best cases, this would mean that after 15 km the probability that the quantum signal reaches the other end is just 50%. Current commercial QKD systems can tolerate around 20-30 dB losses, reaching distances of around 100-150km. Now, in passive optical telecommunications networks, there is not just optical fiber, but also splitters, filters, multiplexers, etc. that introduce additional losses, making things even harder and reducing this distance to essentially metropolitan areas and even limiting them to access segments in some cases. Moreover, if quantum and classical signals are transmitted over the same fiber, other phenomena like four-wave mixing, scattering or optical reflections further degrade the transmission performance of quantum signals. Obviously, the usual electro-optical conversion is not possible since it also destroys the quantum signals, as it actually implies measurement and replication using other media. The losses also affect classical signals and this is why optical amplifiers are a common component in telco networks. Because of the same reason, amplifiers destroy the quantum signals and they must be bypassed. The co-propagation of classical and quantum signals poses yet another problem, since stray photons from a classical pulse appearing in the quantum channel induces a large error rate that quickly makes impossible the successful execution of the quantum communications protocols.

---

[†] i.e. all media except perfect vacuum in which case other issues like aperture will limit the transmission range.

Notwithstanding all these problems, quantum communications is today a viable proposition that is increasingly gaining traction in the market. In the short term as a technology that offers security primitives with unique properties at the physical level and, in the long term, as a way to communicate quantum processing elements, creating the quantum analogue of the current Internet.

In this chapter and for the sake of completeness, we briefly review the basic elements of the quantum communications: qubits and the properties that make them unique information-processing elements and the basic protocols used, for quantum cryptography in the shorter term, and for quantum repeaters in the longer term. We give an idea of what has been achieved and their maturity as telecommunications-ready products and then discuss their integration, at the physical and logical level in the telecommunications infrastructures as well as their application in the industry.

Technological description and state of the art.

When dealing with signals at the quantum level, the fundamental information unit is the qubit. Coined in 1994 by Schumacher and Wooters [19], it is physically embodied by a two-states quantum system. One of the states represents the computational "0" and the other is the "1". Being quantum-mechanical states, their mathematical representation coincides with state vectors in a Hilbert space of dimension two. Technically, they are the solutions of the Schrödinger equation for that particular system. Physically they might be embodied in a myriad of possibilities, like the horizontal/vertical polarization states of a single photon, the spin up/down states of an electron, atom or nucleus, the charge or magnetic flux through a Josephson junction, etc. Each one of them would have a particular mathematical representation -a wave function- from which all physically meaningful quantities of the system can be calculated. From a pure quantum information perspective, their specific mathematical form does not matter and these base states are customarily represented using the so-called Dirac notation as $|0\rangle$ and $|1\rangle$. Since they are the solutions of a complex linear differential equation, any linear combination of these is also a solution. This is the superposition principle, which means that any state belonging to the Hilbert space expanded by $\{|0\rangle,|1\rangle\}$ is also valid and that the general form of a qubit is then a *superposition* $\alpha|0\rangle + \beta|1\rangle$, where the normalization condition $\alpha^2 + \beta^2 = 1$ holds and $\alpha, \beta$ are complex numbers. A direct consequence for information processing is the no-cloning theorem [2] mentioned above. This might also mean that a qubit could store an infinite amount of information in the $\alpha, \beta$ values, however this is not the case due to other of the striking characteristics of quantum mechanics: when a state like the one written above is measured, there are only two possible outcomes: either we obtain $|0\rangle$, with a probability $\alpha^2$, or $|1\rangle$, with a probability $\beta^2$. There is no direct access to the $\alpha$ or $\beta$. The only way would be if we have many copies, in which case we could perform statistics to determine the $\alpha, \beta$ values. When we have a single quantum this is not possible and we only get a $|0\rangle$ or a $|1\rangle$. After the state is measured, it *collapses* to the state corresponding with the result of the measurement, $|0\rangle$ if we got "0" or $|1\rangle$ if we got "1". The probabilities of each outcome given by $\alpha^2$ and $\beta^2$, respectively. It is important to note here the intrinsic *randomness* built in the quantum world, since it makes possible to create random number generator devices

-sources of entropy- rooted in a fundamental law of nature, not in some inability to compute an outcome from a physical process. These properties would allow us to design a protocol to do Quantum Key Distribution, but there is yet another property that is even more striking and that can be used to overcome the distance problem in quantum communications. This property is called *entanglement* and arises naturally when describing states with more than a single quantum in a Hilbert space. The description in Hilbert space of a system with two quanta $|\varphi\rangle = \alpha|0\rangle + \beta|1\rangle$ and $|\phi\rangle = \alpha'|0\rangle + \beta'|1\rangle$ is just their tensor product:

$$|\varphi\rangle \otimes |\phi\rangle = (\alpha|0\rangle + \beta|1\rangle) \otimes (\alpha'|0\rangle + \beta'|1\rangle) =$$
$$= \alpha\alpha'|00\rangle + \alpha\beta'|01\rangle + \beta\alpha'|10\rangle + \beta\beta'|11\rangle$$

Where we have made explicit that now we are in a Hilbert space of dimension 4 with basis vectors $\{|00\rangle, |01\rangle, |10\rangle, |11\rangle\}$. Note that the first index refers to the first qubit and the second to the second qubit. The interesting part is that this shows that the dimension of the Hilbert space grows *exponentially* with the number of qubits and that, by using just two qubits, the Hilbert space contains states like

$$|\Psi\rangle = \frac{1}{\sqrt{2}}(|00\rangle + |11\rangle)$$

The striking feature of these type of states is that they cannot be separated as the tensor product of states of one qubit times the other, i.e. $|\Psi\rangle \neq |\varphi\rangle \otimes |\phi\rangle$, no matter how we choose $|\varphi\rangle$ or $|\phi\rangle$. This non-separability is what characterizes an *entangled state*. Now, if we measure one of the qubits, say the first one, according to the measurement postulate stated above we will find that the second qubit is *always* in the same state. This is, if we measure the first qubit and find that the result is "0", then the second qubit is known to be in the $|0\rangle$ state. A subsequent measurement of this qubit will produce a "0" with certainty. This behavior will happen no matter the distance that separates both qubits and is the source of the "non-classical" correlations underlying the power of quantum information processing. This makes quantum mechanics a non-local theory and is what made Einstein to declare that quantum mechanics was not a complete theory in 1935 [20] and that some form of hidden variables should exist to explain these strange correlations. In 1965 John Bell [21] derived a set of inequalities that allowed to test whether these hidden variables actually could explain the results of certain experiments. The first experiments were performed already in the 70's, but they required conditions extremely difficult to fulfil and it was not till 2015 that a series of three independent experiments in Austria, the Netherlands and the US, confirmed that the quantum theory was correct. Quantum mechanics is arguably one of the most tested and successful theories of physics, and allows for information processing capabilities that are not possible using classical means alone.

In quantum communications the most significant ones are Quantum Key Distribution and quantum teleportation. The first one solves the problem of symmetric key distribution by creating a key that is only known to those executing the protocol at both ends of a quantum channel. The process is based solely on the laws of nature as described by quantum physics. No computational assumption is needed and then the protocol is immune to any attacker, independently of their computational power. This is known as Information Theoretic Security. QKD can limit the amount of information on the key that is leaked to the outside world to any desired level. Obviously, this is true in the mathematical sense. Its implementation in real devices is subject to imperfections, which might reduce its security. As a result, the certification of QKD devices according to its intended security level is an active field of work.

QKD can be performed using a number of protocols. These can be divided in different types considering whether they explicitly use entanglement or not, working in what is known as a prepare and measure protocol, or whether they use discrete or continuous variables. From a security point of view, all of them can be demonstrably secure, however their implementation is very different and have different strengths and weaknesses.

All of them require the ability to produce, manipulate, transmit and measure quantum signals. For telecommunications, these quantum signals are always photons and the physical media to transport them is either the optical fiber or the free space. The degree of freedom used to encode the information is more varied: polarization, phase, phase difference between adjacent pulses, etc. have been used. In the case of entanglement-based protocols [22] [23], the photons need to be created in entangled pairs, which is more difficult and with a lower yield than producing single photons by attenuating a laser pulse. The latter is the easiest one and is used in most of today's systems, while the former can be used to better approximate a true single photon source. This avoids attacks like the photon number splitting (PNS), which can break a QKD system when the pulses have more than one photon. An attacker can, at least in principle, learn which pulses carry more than one photon without disturbing them, block those with only a single photon and let pass the rest while keeping one photon. In this case the attacker can get exactly the same information than the legitimate receiver without the receiver knowing that is being attacked, thus breaking the system. The entangled pairs source can be also located in the middle of the quantum channel, sending photons to the detectors at both ends. This in principle doubles the distance, but the fact that the two photons have to reach the detectors, negates most of the possible advantages. After the signals have been received, the testing of Bell inequalities would determine whether there was an attacker in the line (or too much noise, since the measurement by the environment or by an attacker cannot be distinguished) and if it is possible to extract from the measured signals -through a classical, public, but authenticated communications channel- a final secret key.

It is important to note the need to authenticate the communications of this classical channel. Otherwise, a man in the middle type of attack can be performed. Thus, an authentication method has to be agreed between emitter and receiver. An initial authentication during the installation has to be done, and this is external to the QKD protocol. A number of methods, or combinations of them, can be used, like the procedures followed during the installation of a typical hardware security module. These procedures are well known, and for QKD are required only during the first installation. After this is done, a part of the newly generated key can be used to authenticate the classical public channel in the next round of communications, where a new set of quantum signals are transmitted and a new public and authenticated discussion is carried out to obtain again a new set of secret keys. This authentication is also information theoretically secure [24]. In this way, the authentication of the communications in the classical communications channel connecting the emitter to the receiver in a QKD system can continue essentially forever for all practical purposes.

Because there are also techniques to avoid the PNS attack and the additional difficulties in producing a fast source of entangled pairs, these methods are not being used in practice today. It is likely that they will play a significant role in the future, when quantum repeaters become available. Today's QKD systems use prepare and measure protocols, where a quantum state is prepared by an emitter, sent through the quantum channel and measured by the receiver.

These protocols can be built using discrete or continuous variables (DV, CV). In the first case, qubits encoded in some suitable degree of freedom (polarization, phase…) are used to test whether they have been manipulated or not and then detect an attacker. When a qubit is measured only two values can be obtained, these are the discrete variables.

In the second case, the way to test whether the quantum signals have been modified is somewhat different and does not involve the direct use of qubits encoded in single photons, but the quadratures $\{X, P\}$ of the electric field $X \cos\theta + P \sin\theta$ of the electromagnetic wave that describes them. These are continuous variables and, when they represent a small enough signal, the quantum effects are made accessible and can be used to encode quantum information and also tested for disturbances as in discrete variables. QKD systems built using this mechanism were demonstrated already in 2003 [25]. What makes this method particularly interesting in practice is that detection does not involve the typical single quantum detectors, usually bulky and operating at low temperatures (typically of the order of tens of degrees below 0ºC for Si APDs but near 0ºK in the case of superconducting detectors), but homodyne or heterodyne detection. Albeit this has to be done under very low noise conditions, these methods are essentially extensions of the ones used in classical communications. The expectation here is that CV systemss can be more easily manufactured in large quantities and integrated in optoelectronic components that would drive their costs down. Also, since homodyne detection behaves like a very narrow temporal filter, CV systems are more resistant to noise and co-propagation of the quantum channel together with classical ones in the same fiber is possible [26]. Recent work has shown co-propagation of quantum channel in CV-QKD with one hundred WDM channels [27]. The downside is that CV systems cannot tolerate as much losses as the DV ones. Another issue has to do with the processing of the measured signals to extract the secret key, which is computationally much more intensive for CV than for DV and can actually become a real bottleneck. Current CV systems use FPGAs or GPUs, but to really profit from mass production, either ASICs or VLSI components would be needed, requiring investments that have yet to be done. In market terms, the offer of DV QKD systems appears to be larger, since several companies have this kind of systems in the market or close to production. For CV the offer is more reduced, although several companies are readying their systems and expect to make them commercially available in short.

The transmission of quantum level signals is in itself a hard problem. Even using the optical fiber at its most transparent window and good fibre, losses of more than 0.2 dB/km are usual. If we imagine that we are able to produce single photons on demand at a rate of 1GHz, we will obtain a yield of only one photon per second after about 450 km, one per minute after 540 km… or one per century after about 920 km. Moreover, this photon represents just the raw data, the direct measurements that need further processing to extract the final secret key, with results that depend heavily on the noise. The secret key rate might be orders of magnitude lower than the raw data rate, depending on the circumstances. A further consideration is that passive optical networks introduce losses in other devices that are necessarily used: each connector will add another 0.2 dB, a splitter will add 3 dB each time the signal is divided. A 1:8 splitter will add 9 dB, etc.

Current loss tolerance records for the typical DV and CV protocols are in the range of 40-50 dB and 25-30 dB, respectively. Most of the times, this is enough for a metropolitan area network. However, a new family of protocols, known as Measurement Device Independent (MDI) [28] [29] have been recently proposed that could potentially achieve tolerances in the range of 100 dB reaching distances of 500km in optical fiber, although more pragmatic figures would be around 60 dB and 300km. These systems have also the advantage of having the detectors in a measurement station located in between the path connecting the two emitters. In this case, the emitters act as the end points of the quantum channel: the secret key is actually created at both emitter sites. The detectors can report their measurements openly. Actually, the detectors can be owned by the attacker and this does not make the system less secure. Obviously, the attacker could decide to interrupt the communication by not reporting any measurement. MDI systems are still being developed and no commercial implementation exists.

In any case, there is a maximum range that can be covered by QKD systems and we can safely quote to be around 500 km in optical fiber. Going beyond this using available or near-term technology would require the use of satellites or trusted nodes. The trusted nodes approach consists in the simple concatenation of QKD links whereby the keys obtained in the nodes in the middle of the chain are used to transport (in an ITS way to keep the same security level, e.g. using One Time Pad) the secret key from the origin to the end node. Obviously, the key will be known to the middle nodes, hence the trust. A satellite can be used as a relay in the same way or also an entangled photon source can be placed onboard the satellite to create a key between two ground stations. Proofs of concept have been done with satellites [30] [31] and it appears that this is the only near-term solution for long range QKD without trusted nodes.

In the longer term, it is possible to create quantum repeaters [9]. These are devices that can create long range quantum correlations without actually copying qubits. The basic idea is to use the teleportation algorithm [32]. This algorithm "imprints" the unknown state of an input qubit on a remote qubit, which has to be one of the elements of an entangled pair. This is achieved by doing a joint measurement (which can be seen as two qubit gates) of the input qubit and the other element of the entangled pair. This has the effect of modifying the state of the remote element, making it identical to the original input state, which can be exactly recovered after some transformations selected using classical information obtained during the joint measurement. These manipulations destroy the original qubit while giving no information about it. This effectively teleports the state, which is erased from the input state, such that the no-cloning theorem is respected. The entangled pair acts as a resource that is consumed in the process. The reach of this process is, in principle, unlimited, since the entangled pair can be as separated as we want, but in practice technological limits apply. Since we do not have high capacity and long-term qubit memories, we cannot distribute the entangled pairs beforehand. A naïve way to solve this problem is then to put an entangled pair source in the middle of a length span that has a relatively high probability that both photons are going to reach both ends, say a few kilometers. Then, send one of the photons towards the input qubit that we want to teleport and the other in the opposite direction. When the member of the entangled pair reaches the apparatus with the input qubit, the joint measurement is done and then the state of the input qubit appears in the other member of the entangled pair, which has been travelling in the other direction. In this way, essentially the double of the distance between the source and the apparatus doing the joint measurement is covered. The procedure can then be repeated taking the teleported qubit as input to another teleportation process with another entangled pairs source a few kilometers away. Ideally, this process could be repeated as many times as we wish and then we could teleport states –or entanglement- to any distance. This entanglement could be used as a source to perform QKD with unlimited reach, for example. In practice, this method is extremely difficult, since all the elements in the chain should be synchronized with unbelievable precision. Also there is no guarantee that the first level entanglement has been successful at every hop. Quantum memories are then required, a feat that has not been yet achieved. Albeit great advances have been done in the past few years, quantum repeaters that can be used in telecommunications environments are still in the future.

Despite the lack of maturity of some building blocks for a whole transmission chain, and the limited performances, QKD has already been introduced to transmission networks. Next section will detail the integration of QKD solutions in the telecommunication networks.

Other applications are proposed for quantum networks, which could provide a range of cryptographic functions such as quantum bit commitment [33]; useful for operations such as sealed bid transactions; quantum money [34], useful to prevent double spending; and quantum oblivious transfer [35] [36], which may have useful applications to privacy. However, not all of these protocols offer the same degree of security as QKD. For example, unconditionally secure quantum bit commitment was initially though to be impossible [37], but a protocol has been proposed which is unconditionally secure in the relativistic case, i.e. considering that the finite speed of light makes impossible the information transfer among systems that are out of the light-cone [38]. However, it remains to be seen whether these other new protocols offer sufficient advantage over classical methods to provide motivation for the extra complexity of implementation.

The Integration of QKD in Telecommunications Networks.

Telecommunications networks are comprised of links connecting a number of nodes. These nodes spread from end-user devices to the nodes in the core of the networks. This implies the lengths of the various links can range from several tens of meters to several thousands of kilometers. Considering the limited performances of the QKD solutions, without resorting to trusted nodes or satellites, we will only consider terrestrial applications for the following discussion.

Amongst the telecommunications networks, we can distinguish three segments: the access network, the metropolitan network and the core network. See Figure 1.

The access network covers the distance between the end user and the first concentration point, called the Central Office (CO). Its role is to aggregate the signals from user devices. Currently deployed optical access networks are based on Passive Optical Network (PON) solutions. A typical PON consists of a CO with an OLT (Optical Line Terminal)) linked to ONUs (Optical Network Units) co-located to the residential or business users. A PON architecture is thus a point to multipoint connection and implements passive devices such as splitters with a mutualized feeder fiber up to the first splitter in the ODN (Optical Distribution Network). The splitting ratio is typically 32 to 64 on the drop side, though the standard enables up to 128 users. Bidirectional transmission is performed such that the same fiber transports both downstream and upstream signals. According to the topology, losses can be as high as 28 dB between the end user and the CO.

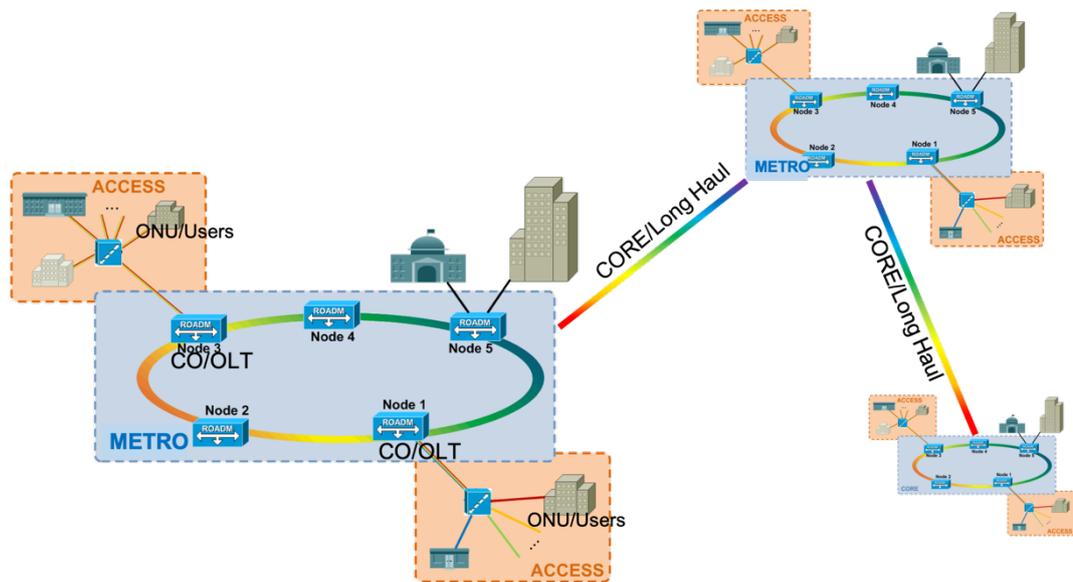

*Figure 1: Simplified view of an optical communications network. A Metropolitan area network links several access networks, where users are connected to the Optical Network Units (ONU). High bandwidth links are required, so typically Wavelength Division Multiplexing technologies are used. Topologies include rings, multi-homed rings and grids. Wavelengths are added/dropped using (Reconfigurable) Optical Add and Drop Modules ((R)OADMS). The access networks aggregate the user traffic at the Optical Line Terminal (OLT) points in the Central Offices (CO). Different metropolitan area networks connect among themselves using the very high-bandwidth long haul connections of the core network, a grid network which heavily use WDM technologies and optical amplifiers.*

The metropolitan network covers the area within large population centers (or among a number of smaller ones) and collects the data from the access networks to send them through the core network. A metropolitan network covers several hundreds of square kilometers and typically uses optically amplified links. The spans between the amplifiers is in the order of 80 to 100 km which means losses of about 20 to 25 dB including connections losses. The links transport, typically using Wavelength Division Multiplexing technology (WDM), several tens of channels in the same fibre, which can transport each from 10 Gb/s to 400 Gb/s data rate each.

The core network relates to the big pipes that transport the traffic between the big population centers of a country (or several ones, European scale for example), and covers several thousands of kilometers. As for metropolitan network, WDM technology is used: 80 to 100 channels in C-band (that can double if using L-band) at 100 Gb/s to 400 Gb/s data rate each, are transported on these links for a total traffic of several tens of Terabits/s.

When considering the integration of quantum technologies in telecommunications networks, there are two different aspects to take into account. One is the integration at the physical level and the other at the logical level. The first one is related more to the nature of the quantum signals and the second one to the products obtained through the transmission and processing of these quantum signals. Given our focus more on short term technologies, we will concentrate on QKD, although part of the discussion is also applicable to general quantum communications.

*Integration at the physical level*

As it has been previously commented, the transmission of quantum signals is a delicate issue. Any interaction with the environment is actually equivalent to a measurement and destroys the information stored in the qubit. As a result, the choice of the degree of freedom used to encode the information is of paramount importance and this has a direct impact on how the quantum device has to be built. For example, the optical fibers typically used in telecommunications do not preserve the polarization, if a qubit is encoded using polarization, a calibration system has to be put in place and strong light pulses have to be sent in order to test the transformation induced and compensate for it. Thus, while polarization is good in birefringent media like the atmosphere, other degrees of freedom, like phase are preferred in optical fiber. Moreover, the need to transmit single-quantum signals, make quantum communications extremely sensitive to losses. As it was mentioned in the introduction to the section, this unavoidably limits the maximum distance (losses) achievable using directly transmitted signals. Fortunately, this problem can be overcome by using quantum repeaters, the devices introduced above that could be used to create quantum correlations no matter the distance. These correlations can then be used to do quantum teleportation, which allows a qubit to be recreated at the end point while being destroyed at the origin, thus preserving the no-cloning theorem.

However, quantum repeaters have yet to be built and demonstrated in real-world networks and, for all practical QKD approaches today, this means that long distances in optical networks can be achieved only using a chain of trusted nodes as introduced before. The existence of these type of nodes must be considered for the current integration of QKD in telecommunications networks.

Due to the difficulty of transmitting and detecting quantum signals, most of the QKD testbeds have been designed as separate networks, avoiding as much as possible the losses and the interaction with classical signals. This situation has been also extended to the control and management structures, that essentially had a minimal connection with the data network where the secret keys are used. It is true that, since the technology has seen a limited deployment, the networks were just a set of point to point connections with pre-fixed key consumers and not much requirements in terms of management. Maintaining a completely separated infrastructure just for QKD and with such a scarce support for connectivity to the data network is a poor proposition, as it incurs in relatively high cost and poor scalability for the services offered.

When a real integration is considered we have to either settle for a design in which a logical infrastructure integrates the management and services of the QKD networks within the existing telecommunications networks, while still keeping a fully separated infrastructure for the physical part, or try to share as much as possible the physical infrastructure, either among many quantum channels [39] or among quantum and classical signals [27, 40] [41]. How much can be shared is limited by physics, but also by the cost/benefit of these approaches.

When physically sharing the infrastructure, some attention needs to be paid to the specifics of the QKD modules generating and detecting the quantum signals and also to the protocols used. In particular, when a pair of modules (emitter/receiver, typically named Alice/Bob) are connected to generate secret keys, three different channels need to be created: a quantum channel to transport the quantum signals, a service channel to monitor an stabilize the quantum channel and quantum signals measurements and a classical channel to carry all the information needed to extract the secret key out of the measurements of the quantum level signals. The last one does not need to share the same physical infrastructure, although it might be convenient from a management point of view. The quantum and service channel need, in many cases, to be implemented using exactly the same physical substrate. Think, for example, on the case previously illustrated where the polarization transformations that affect the qubits need to be monitored. This needs strong pulses with known polarization travelling exactly through the same fiber and also close enough in time. In other cases, this requirement might be somehow relaxed, like when only time synchronization is needed, in which the physical substrate might be different, but still with differences stable enough between the two media (e.g. known length difference and fibers that pass through the same places such that dilation affects them equally). So, if many quantum channels need to share the same physical infrastructure, attention has to be paid not only to the obviously incompatible situations, like when two QKD emitters are installed in the same location, accessing the same physical fiber to implement the quantum channel, using exactly the same wavelength and time slots, but also the way in which the service channel is implemented. This is assuming that the channel implementing the emitter/receiver (Alice/Bob) classical communications does not share the same physical medium than the quantum. Otherwise, this usually implies the need to reduce the power of the classical channel, the use of a band well separated, strong filtering schemes or alternatives that avoid the simultaneous use of the media (e.g. Time Division Multiplexing). When several QKD manufacturers share the same infrastructure, this needs special consideration, since the way that the quantum and service channels are implemented is completely proprietary and this information is not freely available. The further lack of standards dealing with this issue, means that the examples of systems from different QKD manufacturers sharing the same physical media are rare.

The other situation arises when classical and quantum channels share the same physical media. This situation, while desirable from a deployment point of view, has limitations due to the interference that the powerful classical signals generate in the quantum channel. Moreover the coexistence of classical and quantum channels become more critical in bidirectional links for example in point to point Fiber to the Antenna link (FTTA) or point-to-multipoint Access Networks. Bidirectional transmission induces additional penalties arising from reflections and scattering: the penalty originating from these previously ignored noise sources which are at the quantum level have to be considered.

Indeed, several physical effects, like four-wave mixing, Rayleigh backscattering or Raman scattering have the potential to disrupt the quantum channel by flooding it with stray photons. Raman scattering (spontaneous or stimulated) is especially difficult to filter since it is not generated at specific wavelengths, but in a very broad (±200nm) range around the source signal. The most effective way to avoid it, beyond using very narrow filters -temporal or spatial-, is to reduce the power of the classical signals. This is something difficult in some circumstances, like in long-haul lines, that typically uses a DWDM grid heavily populated and with high power and amplifiers to maximize reach. In other situations, like in metropolitan area and dynamical networks, where the power and number of classical channels can be managed, this proposition is much more viable, but still constraining for the operator that has to guarantee the QoS of the transported data. Whether this option is used or not depends very much on multiple external conditions, like the requirements on the QKD services (e.g. key-rate) since it affects very much the performance of the systems, the availability of programmatic interfaces at the low level that permits the dynamic creation of light-paths minimizing the noise or on the availability of fiber for the quantum channel, since in some situations there is simply no other option than to share the physical infrastructure. Finally, new evolutions in the network, like new ultra-low absorption fibers and high sensitivity detectors for classical communications might change the panorama. Certainly, the availability of noise-resistant CV systems or new filtering schemes [41], are promising results in this direction.

*Integration at the logical level*

The quantum communications technologies that implement the fundamental physical capabilities of transmission and detection (and eventually storing) of quantum signals must be integrated into a general network framework aware of these new capabilities and suitable to control them, such that optimal performance can be achieved.

In the past, quantum networks [42] have been deployed as separate infrastructures built ad-hoc and managed in ways that, while suitable for demonstrations, are difficult to scale up to carrier-grade [43] standards. Nowadays, a more global view is emerging that sees the quantum capabilities as an add-on to an existing telecommunications network. A suitable view is to encapsulate these capabilities in a logical structure akin to the data forwarding plane in an SDN network. The advantage to think in this way is that then the integration is reduced to define the interfaces between a classical network and this so-called Quantum Forwarding Plane (QFP). At this level, the functionality and boundary of the QFP is simple to define. The functionality is just to profit from the properties of the quantum states and the capability to transport them. For example, it might be the creation of secret keys or the creation of non-classical correlations (entanglement) among distant quantum states such that other operations can be done (e.g. teleportation of states among memories of quantum computers). Then the QFP ends when this product - the secret keys or the entanglement of distant memories- is created. A clear functionality lets us to define the information flow through the boundary i.e. the interfaces which allow to add the quantum capabilities to an existing network. In the case of secret keys, given their importance in many aspects of the communications, a further network service is usually created on top.

In order to build a quantum/classical network, it is convenient here to introduce a components-based view, since it better conveys the functionality assigned to each one and their relationship, which defines the interfaces. Open interfaces and well-defined functionality is also important for the telecommunications providers and for the broad acceptance of the technology, since it supports technological disaggregation. This means that different manufacturers can provide different components that can be easily integrated in the network through the interfaces. In a security infrastructure this is of particular importance, not only to avoid vendor lock-in, but to guarantee future security and continuity. From a manufacturer point of view, it allows to enrich evolution roadmaps to incorporate new technologies in existing products, overhauling existing know-how with new concepts and perspectives.

In the case of a security infrastructure the key management functionality is of particular interest. A Key Management System (KMS) is a software toolset designated to help in the key management process [17]: the creation, exchange, storage, use and destruction of keys. Note that this is not just a cryptographic algorithm but a set of them together with procedures and policies. As such, it is a complex system and there are many companies that are specialized in KM. QKD brings new capabilities, with keys created following completely different protocols and patterns and require modifications to the existing KMS. This is a possible entry point for new companies in the QKD world, also very beneficial to QKD, since it provides a direct way to integrate QKD into the general security ecosystem. Telecommunications companies use already KMS and for the integration of QKD in their networks, the availability of QKD aware KMS is a must. Ideally while keeping their existing KMS systems.

The KMS is not the only logical component needed for the integration of QKD in telco networks. A minimal set requires also a controller that manages the QKD modules and the routing of the quantum signals. In a network based on fixed links, this module is not needed, but in a reconfigurable and mixed quantum/classical network with infrastructure sharing a rather sophisticated one would be needed. The module not only needs a knowledge of the topology of the network, including the attached devices and their characteristics, but should also have information in real time about dynamical magnitudes like the total power in a given segment and other soft information, like priorities of the service, such that it can calculate whether the instantiation of a quantum channel through a given light-path is physically possible and also meets the conditions imposed by the management rules. Finally, it also has to provide the information needed for accounting and charging for the service when appropriate.

Since, given the current -and expected short term- state of quantum communications technology, it would be impossible to reach beyond a metropolitan/regional area (except if satellites are used), the usage of trusted nodes will be mandatory. This means that, in order to reach beyond the 30 dB loss limit (around 150Km considering very good fibre with 0.2dB/km losses, or 300 km, approx. ~60 dB, if new protocols like Measurement Device Independent or Twin Field [28] [29], today in experimental state, are considered) a trusted nodes model must be considered. This model implies that the secret key from one end to the other, is obtained not in a single quantum jump covering the whole distance that would be impossible, but in a set of smaller jumps where a quantum channel can be created. At the end of each jump a key is extracted. When all the smaller jumps have a key shared between each of them and the next one, then a key can be transferred from the initial to the end point by just consuming each jump's key using a One Time Pad (e.g. XOR-ing the key that we want to transmit with the one used to transport it through that particular jump) It is important to realize that in this model there are two types of keys one that is the effective key that is going to be used by a final application and other that is used just as a transport key. In fact, a QKD network can be envisioned as a mechanism to transport secret keys. As a logical mechanism, it makes no difference if these keys are generated externally to the QKD network (e.g. by a RNG in the initial node) or it is the key that was generated in the, e.g. first QKD link, that is transported to the end point. In any case, the important issue is that in a QKD network there are keys that are going to be used for transport, which are produced only on behalf of the transport function of the network and are never going to be seen by a user application, and others that are the effective final keys consumed by the user application. This distinction justifies the use of another dedicated component: the key forwarding module. It is to be noted that in some demonstrations made up to now, the key forwarding is a task assigned to the key management. Although this is a possible solution, key forwarding/routing is not a typical key management task [17].

The logical components discussed, together with the quantum modules processing the quantum signals, conform a QKD node and is depicted in Figure 2. This architecture can, in principle, fit any of the proposed QKD networks demonstrated up to now [42]. Note that nothing is said about the specific mechanisms for control or key management and both, distributed or centralized architectures can, in principle, be used.

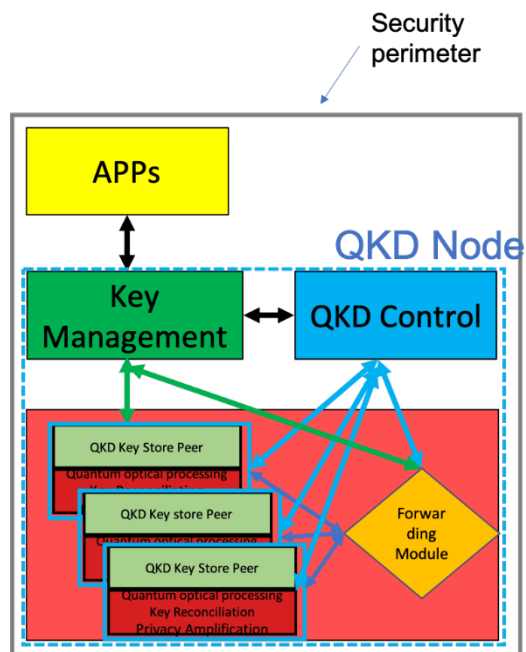

*Figure 2: A generic QKD node with its components and the connections among them. The red part corresponds to the Quantum Forwarding Plane. See text for a detailed description*

These basic architectural concepts, the node and the QFP, also fit well with current networking paradigms, in particular with Software Defined Networking (SDN) [15]. SDN was conceived as a means to add flexibility to the network. The original internet was conceived as distributed by design. The idea was to make it resilient such that it was not possible to break the whole network by just taking down a single node. As such, it was designed as a set of devices and protocols that worked in a mostly autonomous way, without any obvious single point of failure. This meant that the control and data planes were mixed in each node, which in certain aspects took decisions basically on its own. In today's networks this has increased the complexity of the management and make difficult the deployment of new services, that require the modification of many devices, long developing times and lengthy procedures before a new service can be deployed. In SDN, the control and data planes are separated. Also, the data plane is essentially simplified as a data forwarding plane. Network devices are essentially switches, and the network is supervised by a control and management plane which is embodied in a SDN controller. This is just a software instance running on standard computing equipment that communicates with the devices and applications using standard protocols and modelling languages. The resulting architecture very much simplifies the management of the network, making possible a unified Operation Support System, which is very convenient for the telecommunications companies. This also reduces their dependency on a single manufacturer and the deployment of a new service is essentially the deployment of new software, which can be done much more quickly and with much less up-front expenses than in the old model, where new devices needed to be provisioned and installed. This model has also facilitated other paradigms, like the Network Function Virtualization, which is pushing further a transition to a fully softwarized model of the network.

This flexibility makes also possible a much easier and tighter integration of quantum devices in a telecommunications operator network. Under this paradigm is possible to have an information model of a QKD device written in a standard language (e.g. YANG) that is understood by the SDN controller. The QKD system is then just a network device more that is managed according to adequate rules encoded in the SDN controller. This paradigm allows for all possibilities; from a quantum-only network that connects to the classical network in just certain points through a network orchestrator to a fully integrated network sharing a large part of the physical infrastructure. The scheme also allows for an evolutionary upgrade, installing QKD systems only when needed, avoiding large up-front costs. This is in contrast with the old schemes, that required specific modifications node by node and also needed manufacturer dependent modifications, a very difficult task in an emergent market. The SDN-QKD scheme has been recently demonstrated [44]. As a comparison with the generic node, Figure 3 represents the structure of a QKD enabled SDN node. The SDN approach has still other aspect to consider: softwarized infrastructures are acknowledged to be weaker from a security perspective, since a possible weak point can affect the whole system. This problem can be tamed when a physical security layer like QKD is made available to the whole network.

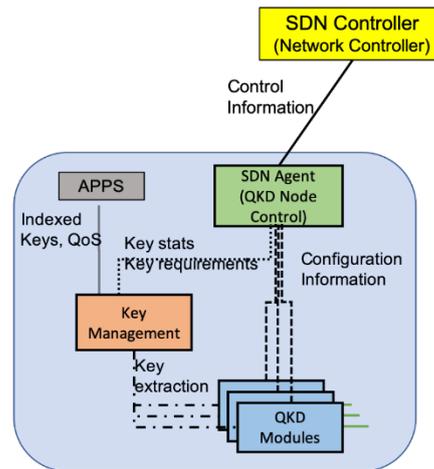

*Figure 3:* Scheme of a Software Defined Networking QKD node showing its main components. The enclosed part corresponds to the node. See *[44]* for a practical implementation

Applications and Adoption drivers.

From a telecom operator's point of view, there are two main parameters which must be differentiated when discussing specific use-cases. The first one stems from the reach limitations of today's QKD technology described in the previous sections. We discern between short-range use-cases, like intra-city data centre interconnection, and long-range use-cases.

We further need to distinguish between 'operator-internal' and 'operator-external' applications. What can be seen as 'operator-internal' application is dedicated to the security of the operator's network itself. This is related to the use of QKD by the operator to protect its own infrastructure (i.e., protection of the control plane or management plane between operator's assets). The wording 'external' refers to the use-cases dealing with the protection of the data of the customers: in that case, the operator can offer protection of the customer data or offer the key as a service (KaaS).

Going into more detail, the first example of a short-reach application is the protection of data centre interconnection (point-to-point application), especially, the traffic exchange between close-by data centres that are acting as disaster recovery backup for each other. Typically, those replacement data centres are geographically disjoint enough to not suffer the same root cause for a failure. Still, they are close enough to each other to avoid effects on the user experience with respect to latency and delay when a re-routing of data becomes necessary. Those distances (20km-80km) allow for a single QKD hop, i.e. there is no need for intermediate trusted nodes on the link between the data centres involved. Also, today's QKD technology is developed enough to achieve sufficiently high secure key rates.

This use-case can further be extended to other applications such as Fibre to the Antenna (FTTA) where a Point of Presence (PoP) or a Cell Site Gateway is linked to the antenna through a fibre (See Figure 4).

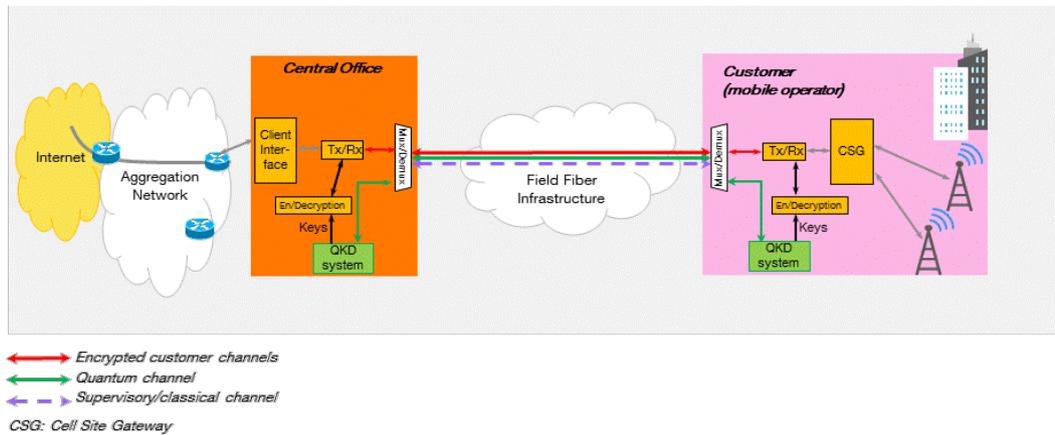

*Figure 4: Point-to-point use-case for mobile backhauling*

Another use-case of interest is in a point-to-multipoint architecture: implementation of QKD in PON in order to secure the transmission of signals between the Central Office to the users. In this case, QKD could be used to transmit securely the keys for the encryption of signals. The main challenge is the high optical losses of a PON system due to the high splitting ratio, although the optical fibre link is relatively short, typically less than 20km. A variation of this use-case is to use multipoint-to-multipoint connection for example linking two clusters of business sites or campuses. In this case, the protection of data becomes mandatory especially in the case of banks, hospitals or governmental bodies.

The protection of own critical infrastructure is a key issue for Telecommunications operators. Operators will protect their network assets like routers, switches, and firewalls by encrypting or authenticating their management and control plane traffic. This kind of traffic is usually found between network operation centres (NOC) and remote network elements (NE). The worst-case scenario is that attackers take on the role of the network administrator by reconstructing the private keys of the certification schemes used, like digital signatures, from the public part. In such a scenario, attackers might change the configuration of domain name servers without even being noticed, because all reconfigurations appear as legitimate changes by an authorized user. This use-case is different to the first one, because network operation centres are usually not close to the administered network elements. The distance between the NOC and the NE might span several hundreds or even thousands of kilometres. A single QKD system is insufficient in such a setup. The solution is obviously, to make use of trusted nodes in between to extend the key generation as needed. Luckily, a network provider owns not only far-away locations, but usually covers a whole area with central offices. They can be used as secure sites and host the appropriate trusted node functionality. Even without spoiling the security model, because the telco operator is its own trust provider and need not fear a breach of its own quantum secure secrets.

While the previous use-cases described so far belong to the class of 'internal protection', the protection of external user data traffic is also an important one. Here, we do find an issue about the usage of trusted nodes: the customers need to rely on the key provider to not misuse the knowledge of the openly accessible keys along the chain of trusted nodes. There are some ways out of the dilemma, though. Currently discussed are ideas like Shamir's secret sharing or to use multi-path approaches in which customers are using different provider networks along disjoint paths and have a separate key management on both (or all) its connected branch offices.

# Quantum Randomness

Randomness generation is a core element of a variety of IT technologies, ranging from security to computing or applications. Indeed, it is also required for QKD, where qubit values and coding basis are assumed to be chosen randomly.

However, producing high-quality random numbers and supplying them at the required speed is a challenging task. The main reason is that randomness can only be generated by measuring a physical random process, and it is therefore a challenge that cannot be solely solved by software. The engineering and production of physical random number generators (a.k.a. true random number generators) has a long history [45]. Remarkably quantum technologies have now matured to the point at which quantum processes can be used to generate the highest possible quality random digits. The randomness derived from quantum processes is a by-product of the fundamental inner workings of the quantum world, in contrast to the randomness derived from classical processes, in which randomness emerges from lack of information or ignorance on the system. The randomness extracted by sampling a quantum process is therefore safer to generate *truly* unpredictable digits. In this section, we introduce the topic of quantum random number generation and its application in the telecommunication industry.

Technological description and state of the art

Quantum random number generators (QRNGs) are devices that produce random digits from the detection of an unpredictable quantum process. The main advantage of QRNGs compared to today's solutions is that the amount of unpredictability that is generated (a.k.a. the amount of available randomness or min-entropy) can be estimated and derived from first principles. This ensures the highest possible quality source for entropy as well as a new class of testing and validation possibilities.

Many different QRNGs schemes can be found in the literature, nearly all of them are based on quantum optical processes of light generation and detection. It is helpful to classify these in different ways, and we will focus on (i) the kind of optical states (few photons or many photons) and (ii) the entropy estimation method (using a full physical model or a simplified physical model plus a challenge-response process).

QRNGs by the type of optical states:

- **Discrete variable approaches.** In this type of device, single photon technologies are used for producing random numbers. Typical approaches include splitting a single photon on a beam splitter [46], measuring the time of arrival of a single photon on a single photon detector [47], or measuring entangled particles [48].

- **Continuous variable approaches.** These devices employ macroscopic states of light containing many photons, and suitable detection technologies. Some examples include measuring shot noise on a CMOS camera [49], phase noise on a CW laser [50], vacuum fluctuations on a homodyne detector [51] or phase diffusion on a pulsed semiconductor laser [52]. All commercial devices today fall within this category, with performance ranging from a few Mb/s to Gb/s and form factors ranging from stand-alone chipsets to appliances.

Another important aspect of QRNGs is the entropy estimation method that is implemented. There are three approaches that are commonly used today:

- **Validated physical model (VPM):** A detailed physical model of the device is developed and validated by detailed characterization of the device components. Critical parameters of the device are measured, and confidence bounds are placed on their possible values. The entropy resulting solely from trusted quantum physical processes is calculated from the model, making worst-case assumptions about device parameters [53] [54].
- **Fully challenge-response protocol (CRP):** Entropy bounds can also be derived by exploiting the unique statistical properties of quantum particles in challenge-response protocols. A typical scenario for such fully CR protocol is a loophole-free Bell test, in which one can derive statistics by processing the input and output data with minimal characterization of the hardware. This allows randomness expansion with no component modelling at all, known as *device-independent* randomness extraction because the entropy estimator refers only to the challenge and response data [48], and not to any device parameters [55].
- **Mixed VPM/challenge-response protocols (VPM/CRP**): Fully CRP schemes require large infrastructure and are therefore hard to implement in practice. A more practical approach uses a mixture of VPM and CRP methods. For some part of the device, a detailed model is developed by VPM, while other components (e.g. the light source) are not explicitly modelled and the security is derived from the challenge-response (input/output) data analysis. There are different proposals in the literature [56] [57] using assumptions as bounded energy or bounded dimensions, among others.

In brief, QRNGs will either be discrete variable or continuous variable with respect to the physical carrier employed and use VPM, CRP, or mixed VPM/CRP methods for the entropy estimation process. It is worth highlighting that in CRP processes, there are as well hardware assumptions that have to be made and tested against.

### Quantum entropy and randomness within the telecommunications industry

In contrast to other quantum technologies, QRNGs can be directly used in today's telecommunication systems. In the most simplified form, QRNGs can replace today's PRNGs, bringing the advanced security capabilities and increased performance that a trusted entropy source guarantees.

The first and simpler integration scheme to interface QRNG devices with IT equipment is to employ standard network interfaces. There are at least two ways to do this. In the first scheme, the QRNG device is plugged "one-on-one" with the target system, as for instance a hardware security module (HSM), via a dedicated Ethernet interface. The second approach is based on making the QRNGs available as a network resource. Examples of this second approach, applied to network management tasks, were used recently for a demonstration of the Ordered Proof of Transit (OPoT) protocol [40], generating the link masks by means of 240 Mb/s of QRNG data to ensure the topology verification of a network service on a 100 Gb/s link. Similar approaches are being described in different domains, following NIST's Entropy-as-a-Service architecture [58].

The second level of integration of QRNGs is through the direct embedding within the IT equipment. In this case, QRNGs become an intrinsic piece of the telecommunications infrastructure, which directly provides advanced security and performance in a scalable and transparent way. In this second approach, the size, power, and scaling of QRNG components are of critical importance. Significant progress has been made recently on the integration of QRNG to meet these requirements [59, 40] [49] [60]

Applications and Adoption drivers

The applications of QRNGs in the information and telecommunications domain is not limited to cybersecurity, and other applications can also profit. We describe here 3 general schemes for the use of QRNGs: (i) increasing security and performance of the infrastructure, (ii) high-performance computation and (iii) applications.

**Increasing security and performance of the infrastructure:** Transitioning into quantum-safe IT architectures is an important endeavour of many organizations today. The term cryptoagility is typically coined to refer to a system design that is flexible to adopt new cryptographic schemes as soon as they become available, with minimal impact on the operations. This is of particular relevance in designing IT systems with a quantum-safe protection capability in mind. Quantum random number generators provide a natural choice for a future-proof design in both quantum and post-quantum cryptography schemes. QRNGs provide higher performance and security features, as well as increased confidence in the unpredictability of this key component. For instance, QRNGs can be used to improve the entropy generation process in IT equipment. This can lead to significant performance and security improvements in cloud environments to mitigate the risk of entropy starvation [61]**.**

**High-performance computation.** Numerical simulations utilize random numbers in so-called Monte Carlo methods and other statistical simulations. There are two opportunities for QRNGs in this domain. First, to improve the results of the simulations compared to those achieved with pseudo-random numbers, which is of relevance in randomness-intensive simulations. It has been long known that some well-regarded pseudo-random generators have produced unexpectedly wrong results in specific Monte Carlo simulations [62]. Using true random numbers can reduce the risks in statistical calculations that are critical. Second, by offloading the production of random digits, the computing resource can be more optimally utilized, reducing computation time and/or energy consumption.

**Applications.** Random numbers can also be used in a variety of applications beyond cryptography and computation, including gaming, gambling, blockchain or decision-making. The output from the QRNGs (or QRNG-seeded systems) can be directly made available to telco customers running processes on the telecommunication infrastructure. This can bring higher performance, security and trust into the overall solution.

## Quantum Computing

Quantum communications deals with quantum states in low dimensional Hilbert spaces; the information processing needs are limited and one or two qubits -in the case of an entangled pair- or just a few of them -in case of multipartite protocols- is all that is needed. This is what makes quantum communications a realistic technology in the short term. However, to reap the benefits of quantum computing, we have to deal with many qubits at a time and this makes the problem far more complex.

Two qubits can be in a superposition of four states, three qubits can be in a superposition of eight states… and so on. Therefore, generalizing while N bit can take one of $2^N$ possible permutations, N qubit can stay in a superposition of all $2^N$ possible permutations. This is very difficult to control, but has remarkable consequences in computation.

A quantum register - associated to N qubits - may have a state which is the superposition of all $2^N$ values simultaneously: therefore, by applying a quantum operation to the quantum register would result in altering all $2^N$ values at the same time. However, strictly speaking the operations that can be carried out on an ideal universal quantum computer are unitary transforms on the quantum state over the Hilbert space, so the idea of parallelism has limitations. This property allows quantum computers to transform qubits with "a sort of parallel computation" reducing the processing time (sometimes dramatically: from exponential to polynomial time) for solving certain complex problems.

In general, there are two main classes of quantum computers: analog and gate-based. Analog quantum computers include annealers, adiabatic computers, i.e., systems which solve problems by directly manipulating the interactions between qubits rather than breaking actions into a set of gate operations.

Gate-based quantum computers, sometimes referred to as universal quantum computers, use logical gate operations (AND, OR, etc.) on qubits. Quantum logic gates are the building blocks of quantum circuits: for example, CNOTs and unitary single qubit operations form a universal set of quantum computing.
It should be mentioned that it is also possible to simulate quantum gate-based computers by using classical computers. There exists a variety of software libraries that can be used, each with different purposes: a comprehensive list of tools is available on Quantiki [63]. Simulation can be made, for instance, using OpenCL (Open Computing Language) [43] which is a general-purpose framework for heterogeneous parallel computing on standard hardware, such as CPUs, GPUs, DSP (Digital Signal Processors) and FPGAs (Field-Programmable Gate Arrays).

There are multiple ways to build gate-based quantum computers manipulating qubits. Table 1 provides an overview (not exhaustive) of the many options available: superconductors and trapped ions are presently the most promising implementations [64].

*Table 1: Examples of gate-based approaches for developing quantum computers*

| Superconducting | Spin | Topological | Ion Trap | Neutral Atoms | Photonics |
|---|---|---|---|---|---|
| Superpositions of currents flowing in superconductors | Qubits encoded in the spin degree of freedom (e.g electrons confined in quantum dots, NV centers, nuclear spin in NMR or impurities embedded in a substrate) | Topological quasi-particles (e.g., Majorana particles) | Ions trapped in electric fields (vacuum and lasers manipulate quantum states) | Atoms trapped in magnetic or optical fields (vacuum and lasers manipulate quantum states) | Qubits encoded in quantum states of photons |
| **Players** | | | | | |
| IBM Rigetti Google Alibaba | Intel Quantum Brilliance SpinQ | Microsoft | IonQ Honeywell AQT | Pasqal ColdQuanta QuERA | Psi Quantum Xanadu ORCA Quandela QuiX |

## Quantum Software

Most of the optimization problems in the field of telecommunications and ICT are currently solved with algorithms finding suboptimal solutions, because of the excessive cost of finding an optimal solution. Examples of these problems includes: joint optimization of several functions, such as radio channel estimation, data detection and synchronization, Data Center optimization, Artificial Intelligence methods, etc.

Quantum computers can help in solving these problems in shorter time or even aiming at optimal solutions. Other applications domains, contiguous to the telecommunications and ICT, includes: Precision Medicine and Biology, Energy, Finance, Smart Cities and Transportation

A list of examples of applications domains for Quantum Technologies, many of them of direct interest for telecommunications, either because they solve problems arising in telecommunications or because they impact in areas that would require the cooperation of telecommunications to be solved, are presented in Table 2

*Table 2: Examples of applications domains for Quantum Technologies*

| Domains | Quantum Communications | Quantum Computing | Quantum Simulation | Quantum Sensing & Metrology |
|---|---|---|---|---|
| **Telecom and ICT** | Quantum safe communication (e.g., QKD, QRNG) | Infrastructure optimization planning and operations; Artificial Intelligence (AI) | Infrastructure simulations: e.g., traffic, energy, resources… | Clock synchronization; more accurate sensors |
| **Industry 4.0** | Quantum safe communication (e.g., QKD, QRNG) | optimization planning and operations; Artificial Intelligence (AI) | Industrial processes simulations Quantum Twin | Automation; more accurate sensors |
| **Precision Medicine and Biology** | Security and protection of patients' data | Improved diagnostics; drug design | Proteomics, Genomics, Drug simulations | Improved sensing for diagnostics imaging |
| **Energy, Oil and Gas** | Security for critical infrastructure | Optimization; Logistics | Predictions and risks analysis | Through-ground imaging |
| **Finance** | Secure transactions | Portfolio management | Portfolio management and trading simulations | Clocks for trade synchronization |
| **Smart Cities and Transport** | Security and data protection | Traffic, resources optimization complexity management | Predictions and risks analysis | Timing synchronization; more accurate sensors; quantum LiDAR |

Today, Quantum annealers (e.g., D-Wave [65] [66]) are already being used to solve some combinatorial and optimization problems. Quantum annealers embody the adiabatic quantum computing model [67], which is formally equivalent to the gate-based model [18]. Nevertheless, current quantum annealers do not present the same level of programmability than gate-based quantum computers: they are specialized computing system tuned to solve optimization problems. In most cases, the problem to be solved is encoded into an Ising-type Hamiltonian, which is then embedded into a quantum hardware graph to be solved by a quantum annealer.

Gate-based quantum computers use another approach. For instance, Figure 5 shows the comparison of the two approaches for the execution of quantum algorithms workflows. In the gate-based approach the problem is formulated in a way for selecting a proper quantum algorithm. Then the quantum algorithm is transformed in a quantum circuit (i.e., using quantum gates) which is either executed on a quantum processor or simulated.

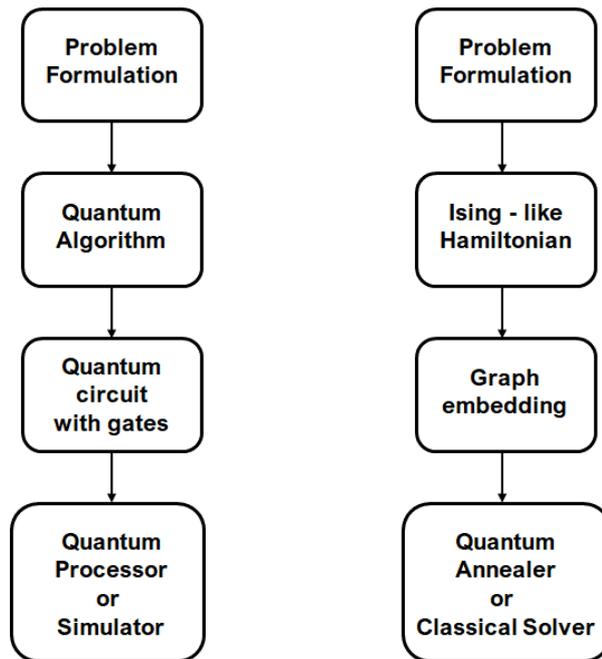

*Figure 5:* Quantum algorithms workflows: gate-model computer (left), quantum annealer (right).

In both cases, random fluctuations (e.g., heat or quantum-mechanical phenomena), could occasionally decohere or randomize the state of qubits: introducing errors and potentially derailing the validity of the calculations. This is why many of these quantum systems require special vacuum environments and the adoption of cryogenic systems. While some algorithms such as Quantum Approximate Optimization Algorithm (QAOA) [68] can tolerate some level of qubit errors, algorithms designed for a Universal quantum computer require logical error corrections methods. However, quantum error correction involves a substantial multiplication of resources: the number of physical qubits required may be orders of magnitude greater than the number of error-free logical qubits seen by the algorithm. In addition, a fairly high fidelity must be realised for any known quantum error correction method to be applied. Because of these problems, many researchers are focusing in the so-called NISQC [69] -Noisy Intermediate Scale Quantum Computation- algorithms. This is the field of study interested in the algorithms and devices that can perform useful computations when the resources are constrained by e.g. a limited number of gates that can be performed before decoherence is too high, a limited number of qubits is available, etc.

Probably, the two best known algorithms are the Shor's algorithm [10], able to factor a number into its prime factors with a superpolinomial gain over the best classical algorithm known, and Grover's algorithm [70], able to search a unstructured database with a polynomial gain over the best classical possible. The first one has been particularly important because it essentially breaks all currently used public key cryptography. The website "Quantum Zoo" [71] has gathered a comprehensive list of algorithms, briefly describing their operation. While quantum computers present a potential threat to cryptography, they also may have positive applications to planning and scheduling problems in telecommunications [72], as well as to acceleration of machine learning.

It should be mentioned that for near term quantum applications, hybrid quantum/classical algorithms are also very promising. A common characteristic of these approaches is that the quantum computer is rather simplified: it is only in charge of carrying out a subroutine, acting as a "coprocessor" while the larger scale algorithm is governed by a classical computer. In this case a higher error rate per operation is tolerable. It may even be possible to implement such quantum algorithms without quantum error correction (as for QAOA [73]).

In summary, when comparing quantum algorithms with their classical counterparts, it appears that employing quantum systems specific performance targets may be reached at a lower computational complexity: on the other hand, an analytical demonstration of the levels of efficiency of quantum computers and algorithms in addressing computational complexity require further studies.

Concerning software languages and tools, the scenario is very active but still rather fragmented: the reference [74] provides an overview of open-source software projects and encourages the coalition of larger communities.

## Quantum Metrology

### Quantum Clock

When we think about clocks, we often think about a wristwatch or the clock in our phones. Those devices have a small internal oscillator, typically made of quartz, that produces a periodic electric signal. This signal is then counted by the electronic circuit to be shown by a display. However, individual oscillators are subject to small variations, so to have our clocks agree with each other we need to synchronize them periodically to a reference. For many years the reference was given by the solar year, but since 1967 the primary reference for the second has been defined by an atomic transition. Nowadays, the massive Cesium fountain and smaller Cesium beam clocks are used to count the time in many measurement institutes around the word. All these measurements are combined to define TAI (Temps Atomique International) then UTC (Coordinated Universal Time).

An atomic clock is a device that uses atoms as reference for the generation of a specific frequency. In the case of Cs, this atomic transition is exactly at 9.192631770 GHz and corresponded to the unperturbed ground-state hyperfine transition frequency of the caesium-133 atoms. This signal is used to lock a quartz oscillator and then lower frequency output signals are synthesized for practical use, typically 5 or 10 MHz and 1 PPS (Pulse Per Second). In practice, several external factors contribute to shifting this frequency when measured.

Using Cs fountain is possible to reach a fractional accuracy of ~$3\times10^{-16}$ [75]. However, Cs clocks are reaching their theoretical limit, and for that reason, in the last 30 years, researchers start to develop new clocks that use transition in the optical domain (>100 THz) that now surpass the performance of Cs clocks (see Figure 6).

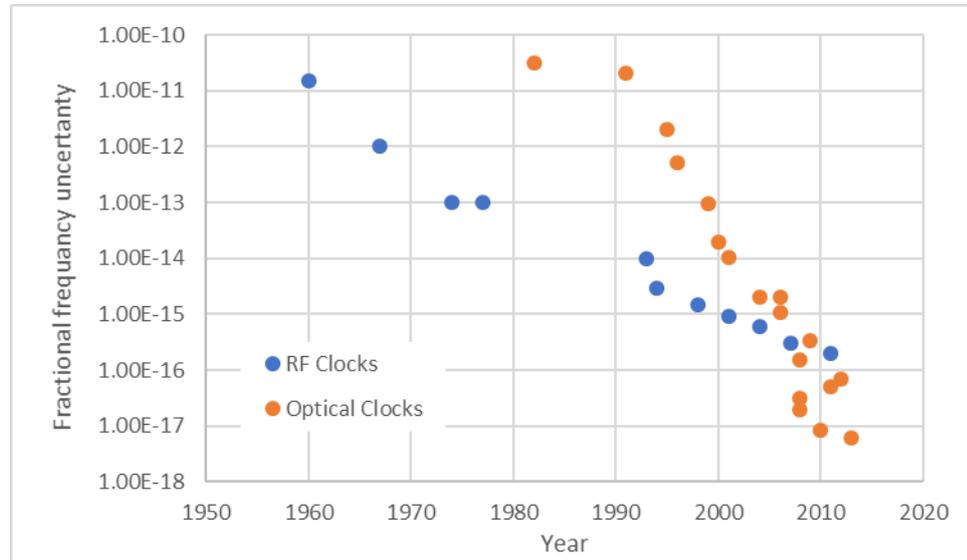

*Figure 6 Improvement of clock stability over the years. Since 2008, optical clocks have outperformed the Cs standard, the uncertainty after this year as been calculated comparing optical clocks among themselves [78].*

By using transition with a frequency tens of thousands times higher than the one used by the Cs standard, the fractional instability of those clocks can reach a much lower level. Modern optical clocks can reach a fractional frequency instability of $10^{-18}$ over 44,000 seconds (e.g. Ytterbium) [76]

In this section, we briefly describe what those new clocks are and how they work, we will describe the state of the art then discuss its application in the telecommunications infrastructures.

Technological description and state of the art.

The most common example of an optical clock is made by a laser that is stabilised in frequency and phase by locking it to an optical cavity[‡] (another example would be ion trap technology). The cavity provides short term stability to the laser. The laser is then used in high-resolution spectroscopy to lock its frequency to the one that corresponds an atomic transition (see Figure 7). The atomic transition used in this spectroscopy has a very narrow linewidth and low sensitivity to external factors like a magnetic field or black body radiation.

---

[‡] *A simple optical cavity is created when two mirrors are facing each other. If a laser with a wavelength that is a multiple of the distance between the mirrors, is injected into the cavity, part of the light will be able to escape the cavity. Otherwise, the light will be absorbed during the multiple bounces between the mirrors. By locking the laser to such cavity, we can then create a relation between the frequency of the laser and the distance between the mirrors. To keep the distance of the mirror constant, those are separated by very stable material like ULE [87] and are kept under vacuum with their temperature stabilised. With the uses of the cavity is possible to obtain lasers with mHz linewidth and $8\ 10^{-17}$ stability [90]*

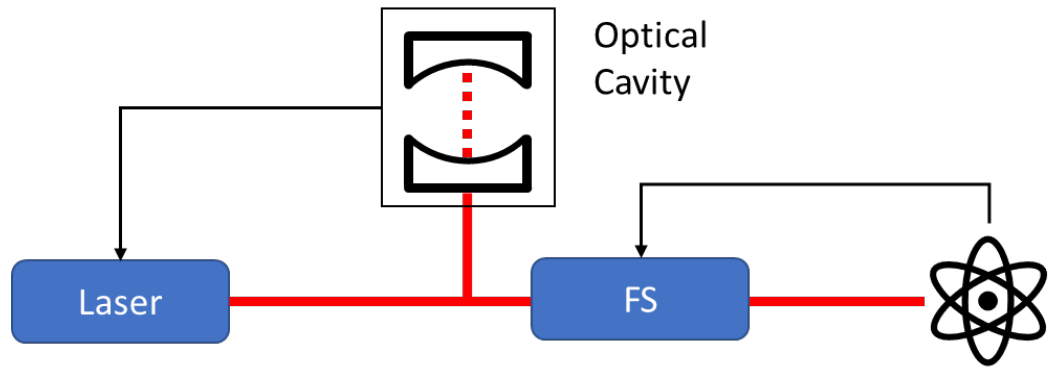

*Figure 7: The buildings block of an atomic optical clock: The light of a laser is sent to an optical cavity and to the atoms. The optical cavity provides short term stability by providing feed back to the laser. The light going to the atom is steered by Frequency Shifter (FS) that is regulated by a resonance of the atoms themselves.*

Two new competing quantum technologies are now being developed: lattice clocks and ions clocks. The first use a cloud of neutral atoms trapped in an optical lattice, the later uses a single ion or multiple trapped ions.

Lattice clocks usually have a better stability thanks to the larger number of atoms interrogated during the clock sequence. Ion clocks have better accuracy since it is easier to isolate them from external perturbations [77] [78].

Once the laser has been stabilised, it needs to be converted to the electrical domain to be used by other equipment. This is done by locking an optical frequency comb to the laser [79]. An optical frequency comb can transfer the stability of the laser (few 100s of THz) to an RF signal (~1GHz) or alternately change the wavelength of the laser used in the clock (UV-visible) to one more suitable for the transmission via fibre (C-band).

Quantum Clocks in the Telecommunications Industry.

Digital communication requires some synchronisation between transmitter and receiver. If the clocks time and phase of the transmitted and received signal do not match, we can incur an error that can corrupt the message or make the transmission less efficient. This means that every electronic device that needs to transmit or receive data need to have its internal clock able to synchronise with its counterpart. When more than two devices are connected in a network, a global synchronisation is required to avoid a bottleneck. Modern networks reach this global implementation by using the Precision Time Protocol (IEEE 1588-2008 and in the future 1588-2019) [80] and/or SyncE (ITU-T Recommendations G.8261 and G.8262). In those standards, a hierarchy of clocks that disseminate time and frequency over the network is defined. On top of this hierarchy sits a master clock that acts as a reference for the network. Those master clocks are usually Cesium clocks that are the most accurate timekeeper commercially available, or is provided by satellite. Thanks to those protocols is possible to synchronise distant location within few tens of ns and with 1.5 µs end-to-end requirement from top of the time synchronization architecture to end equipments. This is sufficient for the current generation of telecommunication; however, we are reaching the limit of this technology. Optical clocks alongside a very specific fibre network (with very special equipments regularly arranged to maintain the stability of the link) capable of disseminate time and frequency without loss of stability, will enable new technology that are not feasible today such as Distributed MIMO and radar, Geodesy, QKD network synchronisation (but other solutions may be available in the future -as High Accuracy profile in PTP 2.1 version- without need of optical clocks networks), Quantum sensing and Quantum network.

Applications and Adoption drivers.

Optical clocks have already surpassed the performance of Cs in the lab and are being considered as substitutes for the new definition of the second. Several projects have been started in the last few years to bring those clocks out of the labs: In 2018 a transportable lattice clock was used to determine the gravity potential difference between the middle of a mountain and a location 90 km apart [81], this year a pair of transportable clocks were used to test general relativity [82]. The iqClock project is pushing this even farther by developing the first commercial optical atomic clock and simplifying its use by the use of superradiant lasers [83] and Microchip Technology is developing a miniaturised ion clock [84]. However, no commercial optical atomic clock exists at the moment, and their development is mostly driven by universities or collaborations via publicly funded projects. More technological development is required to increase the reliability of those clocks and make them ready for deployment. Also, to make use of the full potential of those devices, a network able to transfer the stability of the clock to a remote location will need to be put in place.

## Conclusion and remarks.

In this paper we have reviewed the main applications of the new generation of quantum technologies in the telecommunications industry. In the short term, it is the ability to secure communications links using QKD the one that is expected to have a larger impact. Its maturity is enough to be deployed in the field, although the technology has still to evolve. In particular, the limited range is a major handicap. Beyond metro-area networks or where losses are too high, trusted nodes are required. In these nodes, the quantum mechanical properties that protect the key cannot be used and conventional means have to be used instead. For some use-cases this is of limited importance, as when the trusted node is in a security perimeter owned by the same Telecommunications company that benefits from the use-case, but this might not be the same for others. There are mechanisms to alleviate this problem, ranging from the use of novel protocols, with greater reach and reduced risks in the detectors, to the use of satellites. The ultimate goal is to fully avoid trusted nodes in a global world-wide network. This will happen when quantum repeaters are available, an active research field but still very much in the research stage. A further issue with these technologies is the deployment of systems that are so extremely sensitive to noise in the fibre. This makes problematic to share the quantum channel in a lit fiber with many other classical channels, since the noise coming from the other wavelengths will mask completely the quantum signal. However, the advances have been quick, and new protocols, network paradigms and filtering technologies are making possible to share the existing infrastructure for both, quantum and classical communications. Also, the cost of the technology and its integration with the existing security ecosystem, including security certifications, is steadily improving, helping in bringing what once was an exotic technology to a broad market. QKD is, however, just one of the security-related quantum technologies, and other protocols for tasks like quantum multiparty secret sharing, blind computing, signatures, etc. which are now in the research stage will be made possible with a network that can transmit quantum signals.

Implicit in the QKD protocols is the assumption of the availability of large quantities of random numbers. The true, non-deterministic, randomness afforded by quantum mechanics at a fundamental level is unique to QRNGs, which is different from other RNGs produced because of the limited knowledge of an internal state that can, at least in principle, be known. QRNGs are commercial devices that can be produced at large scale and in small integrated packages, ready to be used in electronic circuits. They have been implemented in mobile phones and are the method of choice for QKD devices. As a reliable entropy source, they have a role in security systems, but also for other applications, from gaming to statistical simulations.

While quantum computing is still in its first steps, it is attracting much interest and large investments from governments and companies. Its future impact in the Telecommunications sector will likely come from two fronts. On the one hand, its new capabilities, especially to solve optimization problems, have many applications in Telecommunications like infrastructure optimization, operations planning, path calculations or even in AI. On the other hand, much like today's computers have reached a far greater functionality because they are connected, quantum computers will benefit from being connected at the quantum level. Thus, quantum communications networks will be also fundamental to maximize the benefits of quantum computers. This will be very important also at the beginning, since quantum computers will mostly work as specialized devices, collaborating with classical computers in accelerating the calculations.

In the field of Quantum metrology and sensing, the advances in quantum clocks and ultra-precise time distribution are the ones more important in telecommunications. Optical quantum clocks are still far from reaching the commercial readiness of other quantum technologies. However, important steps forward have been made, and private companies are starting to gain interest in their development. They will enable pure science-based applications, like tests of unified theories and fundamental constants, large base radio astronomy and dark matter detectors. But they will also push next-generation telecommunication technology as distributed MIMO and quantum networks, with the crucial point of being able to disseminate very high quality signals at the lowest cost.

Finally, there will be other aspects of telecommunications that are very difficult to foresee now that will be impacted by the new generation of quantum technologies. As an example, it is expected that quantum computing and simulation will help in the discovery of new materials and processes that might have large impact in telecommunications, like ultra-low losses fibres, extremely low power signal detection, etc.

## List of Abbreviations

APD: Avalanche Photodiode
ASIC: Application-Specific Integrated Circuit

CMOS: Complementary Metal-Oxide-Semiconductor

CO: Central Office
CPU: Central Processing Unit
CRP: Challenge-Response Protocol
CV-QKD: Continuous-Variables Quantum Key Distribution
CW: Continuous-Wave
DV-QKD: Discrete-Variables Quantum Key Distribution
DWDM: Dense Wavelength Division Multiplexing
DSP: Digital Signal Processor
ETSI: European Telecommunications Standards Institute
FPGA: Field-Programmable Gate Array
FTTA: Fibre To The Antenna
GPU: Graphics Processing Unit
HSM: Hardware Security Module
ICT: Information and Communications Technology
IEEE: Institute of Electrical and Electronics Engineers
IT: Information Technology
ITS: Information-Theoretic Security
ITU-T: International Telecommunication Union – Telecommunication Standardization sector
KM, KMS: Key Manager, Key Management System
MDI: Measurement Device-Independent
MIMO: Multiple-Input Multiple-Output
NE: Network Element
NISQC: Noisy Intermediate-Scale Quantum Computing
NIST: National Institute of Standards and Technology
NOC: Network Operation Center
ODN: Optical Data Network
OLT: Optical Line Termination
ONU: Optical Network Unit
OpenCL: Open Computing Language
OPoT: Ordered Proof of Transit
PNS: Photon Number Splitting
PON: Passive Optical Network
PoP: Point of Presence
PPS: Pulses Per Second
PRNG: Physical Random Number Generator
QAOA: Quantum Approximate Optimization Algorithm
QFP: Quantum Forwarding Plane
QKD: Quantum Key Distribution
QRNG: Quantum Random Number Generator
RF: Radio-Frequency
RNG: Random Number Generator
ROADM: Reconfigurable Optical Add-Drop Module
SDN: Software Defined Networking
TAI: Temps Atomique International - International Atomic Time
UTC: Coordinated Universal Time
UV: Ultraviolet
VLSI: Very Large-Scale Integration
VPM: Validated Physical Model

WDM: Wavelength Division-Multiplexing
YANG: Yet Another Next Generation

# **Declarations**

Availability of supporting data

No additional data is needed for this work

Competing interests

There are no competing interests

Funding

This work has been partially supported from the European Union's Horizon 2020 research and innovation Quantum Flagship projects CiViQ (Grant No. 820466), QRANGE (Grant No. 820405) as well as the project OpenQKD (Grant No. 857156) and project Quantum Information Technologies Madrid, QUITEMAD-CM P2018/TCS-4342, Comunidad Autonoma de Madrid, Spain.

Authors' contributions

V.M. and D.L. conceived the work. V.M. wrote the introductions, edited the paper and, together with J.P.B, C.E. and D.L prepared the structure and contributed to the Quantum Communications section. C.A. contributed mostly to the quantum random number generators section. M.M and O.L.M to the Quantum clocks part. A.M to the quantum computing section. The rest of the authors contributed mainly to the Quantum Communications section. All authors read the paper and contributed to the general readability of the whole paper.

Acknowledgements

N/A